\documentstyle[12pt]{article}
\def\be{\nopagebreak[3]\begin{equation}}
\def\ee{\end{equation}}
\def\ba{\nopagebreak[3]\begin{eqnarray}}
\def\ea{\end{eqnarray}}

	\def\ni{\noindent}
	
\begin{document}
	\begin{center}
	\centerline{}
	\vspace{.5cm}
	{\large\bf Probing Quantum General Relativity through Exactly 
	Soluble Midi-Superspaces II: Polarized Gowdy Models\\}
	\vspace{0.7cm}
	{\large\em M. Pierri\footnote{pierri@gannon.edu}\\}
	\vspace{0.5cm}
	{Department of Physics, Gannon University  \\
	Erie, PA 16541\\ }
	\vspace{0.3cm}
	{\small December 2000\\ }
	\vspace{0.5cm}
	\end{center}

\begin{abstract}

Canonical quantization of the polarized Gowdy midi-superspace with a
3-torus spatial topology is carried out. As in an earlier work on the
Einstein-Rosen cylindrical waves, symmetry reduction is used to cast
the original problem in 4-dimensional space-times to a 3-dimensional
setting. To our knowledge, this is the first complete, systematic 
treatment of the Gowdy model in the geometrodynamical setting. 

\end{abstract}

\section{Introduction}\label{i}

In the canonical approach, general relativity is described by a
suitable set of phase space variables subject to constraints. The
topology of the space-time manifold is fixed once and for all to be
$\Sigma \times {\rm R}$, and the basic phase space variables are
fields on $\Sigma$. Examples of such settings are provided by
geometrodynamics, where the phase space variables are the spatial
metric of $\Sigma$ and its conjugate momentum (essentially the
extrinsic curvature) and the Ashtekar formulation, where the set of
variables is given by a self-dual connection and a (density weighted)
triad.  Quantization is accomplished by constructing a Hilbert space
of quantum states with an appropriate inner-product such that the
phase space variables are promoted to well-defined operators. However,
the detailed procedure faces two technical problems. First, general
relativity has an infinite number of degrees of freedom. Second, there
are non trivial constraints, which are difficult to deal with in the
quantum theory. Indeed, the nature of constraints is such that both
the procedure of solving the constraints classically prior to
quantization and the method of imposing them as operator conditions on
states a la Dirac have proven to be difficult to implement.

To gain insight into some of these issues, a number of models have
been discussed in the literature. Typically, they are obtained from
3+1-dimensional general relativity by symmetry reductions. Perhaps the
simplest example is Bianchi models, where spatial homogeneity forces
the system to have only a finite number of degrees of freedom.  They
have proved to be useful to address technical and conceptual questions
concerning quantum cosmology \cite{at93}.  However, since they have
only a finite number of degrees of freedom, they are not suitable for
investigation of issues concerning the field aspect of gravity.  To
face this difficulty, it is natural to consider symmetry reductions
which are mild enough to leave behind \textit{local} degrees of
freedom.  Spacetimes with two commuting, space-like Killing vectors
fall in this category.  In this case, 3+1 dimensional Einstein's
equations reduce to a system of equations for a two dimensional field
theory.  It has been investigated at the classical level \cite{g71,
av98} and at the quantum level using the Ashtekar formalism in
\cite{g97}.

The next obvious restriction is to assume that the Killing vectors are
hypersurface orthogonal. Then the situation simplifies dramatically.
For, now, in effect, all the non-linearities go away. One can gauge
fix the system in such a way that only one of Einstein's equation has
to be solved, which, furthermore, is equivalent to a
\textit{non-interacting} scalar field propagating on a \textit{flat}
2+1 dimensional space-time.  Since this situation is simple enough to
be exactly soluble, it provides a concrete arena to examine the issues
of quantum gravity and to see how they can be resolved in practice.

In the spatially non-compact case with $R\times U(1)$ isometry group,
such space-times were rigorously quantized in \cite{am96} following
the pioneer work on the subject by Kucha{\v{r}}\cite{kk71} in 1971 and
later work by Allen \cite{ma87}.  They represent (one polarization)
cylindrical gravitational waves, first discovered by Einstein and
Rosen. By performing an analysis similar to \cite{am96}, spacetimes
with topology $R^2 \times {\rm T}^2$ and toroidal symmetries were
quantized in \cite{b98}.

A natural analog of Einstein-Rosen waves but with compact spatial
topology is obtained by taking $U(1)\times U(1)$ isometry group acting
on a compact manifold $\Sigma={\rm T}^3$.  Such space-times were
initially considered by Gowdy \cite{g71} and later canonically
quantized by Berger \cite{b74}. In the symmetry reduction language
used above, the effective linear scalar field now propagates on a flat
but expanding space-time. In \cite{b74} this field was quantized a la
Fock at each instant of time of the flat background metric and the
emphasis was on the resulting ``particle creation''
phenomenon. However, these ``particles'' do not have a natural
physical significance. Indeed, since physically one is dealing here
with a closed system --the symmetry reduced gravitational field--- one
would expect the \textit{physical} vacuum to be stable; there is no
physical external field to pump energy into the system and create
physical quanta/particles.

In this paper, our purpose is to provide an alternative quantization
of the 1-polarization Gowdy models which, we believe, is appropriate
for the physics of the problem. In particular, there will be a
\textit{single} Hilbert space and no particle creation.  We will use
the resulting theory to analyze the conceptual and technical problems
of quantum gravity.  Our analysis is closely related to that of the
Einstein-Rosen waves \cite{am96} and  at appropriate points, we will
compare and contrast the technical steps in the two constructions
and conceptual meaning of the final rsults.

The outline of the paper is as follows.  In section \ref{ghf}, we will
obtain the reduced phase-space for the system from a 2+1-dimensional
perspective. We will see that, before gauge fixing, the two models
(Gowdy and Einstein-Rosen) have formally the same Lagrangian.  But,
the topological differences in the isometry groups and in the spatial
slices introduce important technical and conceptual changes.  For
instance, because of the absence of boundary terms in the action in
the spatially compact Gowdy space-times, there is no longer a
generator of dynamics on the constraint surface. Therefore, in
contrast to the analysis of Einstein-Rosen waves of \cite{am96}, we
will now have to introduce a `deparametrization' procedure to discuss
dynamics. Also, unlike Einstein-Rosen waves, the Gowdy space-times
have an initial singularity. Finally, even after deparametrization, we
will find that a new global constraint and a global degree of freedom
is left over, to be carried on to the quantum theory. The model is
quantized in section \ref{gqt}. We will see that there is a `natural'
Hilbert space for the reduced model. The physical Hilbert space will
be the subspace of the Hilbert space corresponding to the kernel of
the global constraint. In the quantum theory we will investigate some
important issues concerning quantum geometry and coherent states.
Finally, in section \ref{gdiscussion} we will summarize the main
results and point out open questions.

\section{Hamiltonian Formulation}
\label{ghf}

\subsection{The Gowdy system from a 2+1-dimensional perspective}\label{gms}

The 3+1-dimensional Gowdy spacetimes considered here have two commuting hypersurface orthogonal space-like Killing fields.  This system is equivalent to axi-symmetric solutions of 2+1-dimensional general relativity coupled to a zero rest mass scalar field (which is given by the logarithm of the norm of the Killing field).  
Let us begin by specifying our midi-superspace from the 2+1
perspective. 
Thus, we will consider solutions of 2+1-dimensional
general relativity with $U(1)$ isometry group coupled to zero rest
mass scalar-fields (where the $U(1)$ Killing field is hypersurface
orthogonal). The underlying manifold $M$ will be topologically ${\rm
T}^2\times {\rm R}$ and the space-time metric will have signature
(--,+,+).

Denote by $\sigma^a$ the  Killing field. Hypersurface
orthogonality of $\sigma^a$ implies that the space-time metric
$g_{ab}$ has the form:
\be
g_{ab} = h_{ab} + \tau^2 \nabla_a \sigma \nabla_b \sigma
\label{g0.1}
\ee
where  $\tau$ is the norm of the Killing vector field and $\sigma$  
is an angular coordinate with range $0 \leq \sigma <2\pi$   
such that $\sigma^a \nabla_a \sigma =1$.
The field $h_{ab}$ so defined is a
metric of signature (--,+) on the 2-manifolds orthogonal to
$\sigma^a$. 
Let us introduce a generic slicing by compact space-like
hypersurfaces  labelled by $t= {\rm const}$ and a dynamical vector field 
$t^a = Nn^a+N^\theta \hat{\theta}^a $ where $n^a$ is a unit normal 
to the slices and $\hat{\theta}^a$ is the unit vector field within 
each slice orthogonal to $\sigma^a$. The pair $N$, $N^\theta$ 
constitute the lapse and 
shift. If we now introduce an angular coordinate such that 
$\hat{\theta}^a \nabla_a \theta=1$ and $0 \leq\theta <2\pi$.  
Then, the $2$-metric can be written as:
\be
h_{ab} = (-N^2 + N^\theta N_\theta ) \nabla_a t \nabla_b t
	+ 2 N_\theta \nabla_{(a} t \nabla_{b)}\theta + 
	 e^\gamma \nabla_a \theta \nabla_b \theta,
\label{g0.2}
\ee
where $N$, $N^\theta$ and $\gamma$ are functions of $\theta$ and $t$. 
It is because of the underlying $U(1)$ symmetry that the $3$-metric $g_{ab}$ has 
only four independent components and they are functions only of 
two variables.  Moreover due to compactness of the spatial 
slice they are periodic functions of $\theta$.

Thus, our midi-superspace consists of  five functions, 
$(N, N^\theta, \gamma, \tau, \psi)$ of $t$ and 
$\theta$, (which are periodic in $\theta$), 
where $\psi$ is the zero rest mass scalar field. These five fields 
are subject to the following field equations:
\be
G_{ab} =  T_{ab}, \quad{\rm and}\quad 
g^{ab}\nabla_a\nabla_b \psi = 0\, ,
\ee
where $G_{ab}$ is the Einstein tensor of $g_{ab}$ which is determined by
the fields $(N, N^\theta, \gamma, R)$ via (\ref{g0.1}) and (\ref{g0.2})
and $T_{ab}$ is the stress-energy
tensor of the scalar field:

\be
T_{ab} = \nabla_a \psi \nabla_b \psi - {\textstyle{1\over 2}} (g^{cd}
\nabla_c \psi \nabla_d \psi) g_{ab}\,. 
\ee

Note that $T_{ab}$ satisfies the {\it strong energy condition}, i.e.,
$T_{ab}\lambda^a \lambda^b \geq -\frac{1}{2} T$, where $\lambda^a$ is
any unit time-like vector field and $T=T^a_a$.  Now due to compactness
of the spatial slice and the {\it strong energy condition}, the
Hawking-Penrose theorems tell us that the space-time described above
will generically have a singularity. This is a major difference from
Einstein-Rosen cylindrical waves.  There the spatial slices were
asymptotically flat and at the classical level we were mainly
concerned with an appropriate description of the asymptotic structure.
Another difference related to compactness of the spatial slices is
that now $\nabla_a \tau$ has to be time-like on $M$ \cite{c90} whereas
on the non-compact case it is space-like (it is for this reason that
at Ref. \cite{am96} $\tau$ is denoted by $R$).

Before we conclude this section we would like to comment on the
coordinatization of the midi-superspace used here. Let us
look at this system from a $3+1$ perspective.  In this
case the midi-superspace $(\gamma, \tau,\psi)$ refers to vacuum
general relativity. There exists in the literature coordinatizations
for this model which differ somewhat from ours \cite{g71, b74}.  But
the route to quantization is not as direct as the one that we will
obtain here. More details of this comparison will be given later in section \ref{rps}.

\subsection{Canonical Form of the Action}\label{gcfa}

Let us begin with the 3-dimensional action:

\be 
S(g, \psi) := {1\over 2\pi } \int_{M} d^3x \sqrt{g}[ {\cal R} -  
 g^{ab}\nabla_a \psi\nabla_b \psi], 
\label{g5}
\ee
where ${\cal R}$ is  the scalar curvature of $g$.  There are no 
boundary terms because the spatial slice is compact. To pass 
to the Hamiltonian formulation, one performs a
2+1-decomposition. Let us substitute the form of the metric given by  
Eqs. (\ref{g0.1}) and (\ref{g0.2}) in  (\ref{g5}).  Then, the action 
reduces to the standard form
\be
S = \int dt\, \left( \int d\theta (p_\gamma \dot{\gamma} + p_\tau \dot{\tau}
+ p_\psi \dot{\psi}   )
\,\, - H [N, N^\theta]\right),
\label{g5a}
\ee
The Hamiltonian $H$ is given by:
\be
H [N, N^\theta] = \int d\theta (NC + N^\theta C_\theta) 
\label{g6a} 
\ee
where $C$ and $C_\theta$ are functions of the canonical variables:
\ba
C & = &  e^{-\gamma / 2} (2\tau'' - {\gamma}'\tau' - p_{\gamma} p_\tau ) + 
 \tau \, e^{-\gamma/2} (\frac{{p_\psi}^2}{4\tau^2} + {{\psi}'}^2),   
      \nonumber\\
C^\theta & = &  e^{-\gamma} ( -2 {p'}_{\gamma} + \gamma' p_{\gamma} + \tau'p_\tau ) 
+ e^{-\gamma} p_{\psi} \psi' .
\label{g7}
\ea
(Here primes denote derivatives with respect to $\theta$.) 

As expected the lapse and shift $N$, $N^\theta$ appear as 
Lagrange multipliers; they are not dynamical variables. Thus, 
the phase-space $\Gamma$ consists of three 
canonically-conjugate pairs of 
periodic functions of $\theta$, $(\gamma, p_\gamma; \tau, p_\tau; \psi, p_\psi)$ 
on a $2$-manifold $\Sigma$ which is topologically ${\rm T}^2 $.
By varying the  action with respect to lapse and shift we 
obtain as usual two first class constraints $C=0$ and $C^\theta=0$. 
(Because of the underlying symmetry of the  
space-time, the $\sigma$ component of the diffeomorphism 
constraint $ C_\sigma$ vanishes identically.)
Therefore, the Hamiltonian of the system  vanishes
on the constraint surface. 

Note that, because $\Sigma$ is compact, now there is no distinction
between gauge and dynamics. The situation is different from the one we
encountered in the Einstein-Rosen waves. There, the space-time is
asymptotically flat and a non-vanishing Hamiltonian generates
dynamics, whence the standard gauge fixing procedure is well-suited.
In Gowdy space-times, by contrast, $\Sigma$ is spatially compact and
the Hamiltonian vanishes on the constraint surface. Hence one needs to
deparametrize the theory, i.e., it is necessary to select a variable
or combination of variables on the phase-space to play the role of
time. In this sense the compact case is conceptually more subtle and
technically more involved.

\subsection{Deparametrization}\label{gd}

Deparametrization is accomplished by imposing `gauge conditions' on
the phase-space variables such that they extract one point from each
orbit of the Hamiltonian vector field corresponding to the
constraints, \textit{except for one}.  This remaining orbit will
describe evolution.  In another words, from the infinite set of vector
fields generated by the Hamiltonian contraints, we have to select one
to represent evolution and gauge fix the others. For gauge fixing, we
will choose coordinate conditions to make the space-time geometry
transparent. Let us demand:
\be
\tau'(\theta)= 0 \quad{\rm and} \quad   p_\gamma  = - p ,
\label{g11}
\ee
\noindent
where $p$ is a spatial constant. The first condition  will allow us to 
regard $\tau(\theta)$ as the time 
parameter and is motivated by the fact,  noted 
in subsection \ref{gms}, that the field equations imply that 
$\nabla_a \tau$ is  time-like everywhere in $M$. 
The second   condition  states that the scalar density
$p_\gamma$ should equal  $-p$, a spatial constant, in our chart $(\theta, \sigma)$ on $\Sigma$.    
It is equivalent to Fourier analyzing $p_\gamma$ in space and setting all the nonconstant modes %%@
in this expansion to zero.
It will serve to  remove all nonconstant modes of $\gamma$ from our list of dynamical variables.
Let us comment on the presence of this global degree of freedom. The spatial constant $p$ can be %%@
expressed as a function on the phase space by integrating the gauge condition on $p_\gamma$ over 
the circle, namely $p=-\frac{1}{2\pi}\int_0^{2\pi}p_\gamma\,d\theta $. Now we can easily verify %%@
that $p$ has zero Poisson bracket with all the constraints, i.e., it is a Dirac observable.
Therefore it can not be removed by gauge fixing. Thus, if these
choices are admissible, apart from a global degree of freedom, the true degrees of freedom will %%@
all reside in the field $\psi$, in accordance with our general expectation that
in 2+1 dimensions, all the local degrees of freedom are carried by
matter fields.  We should add that the conditions (\ref{g11}) were 
also motivated from  the ones adopted  previously for Einstein-Rosen waves. A 
comparison shows that $\tau(\theta)$ is formally replacing 
the function $R(r)$ in the action.    However their roles are quite 
different due to the restriction on their gradients. 
Just as we chose $R(r)$ as a radial 
coordinate, now we choose $\tau(\theta)$ as a time coordinate. Also, in the ER case we set $p=0$,  
whereas here, as we will see later on, we are not allowed to do so. This is the source of the 
global degree of freedom that is not present on the ER case.
So, for all the similarities,  we might expect to achieve the same technical simplifications  on 
this model.

To see if the coordinate condition is acceptable we have to show   
that the Poisson brackets $\{\tau'(\theta) , H[N,N^\theta]\}$ and  
$\{ p_\gamma+p , H[N,N^\theta]\}$ vanish for a unique choice of 
lapse and shift, or  equivalently   $\dot{\tau}= 1$ and 
$\dot{p}_\gamma=0$. Explicitly,

\ba
\dot{\tau}(\theta) = \{\tau, H[N,N^\theta]\}  &=&  -Ne^{-\gamma/2}p_\gamma + 
	    N^\theta e^{-\gamma}\tau'
	    =1\nonumber\\
\dot{p}_\gamma(\theta) = \{ p_\gamma , H[N,N^\theta]\}  &=&    
	\frac{1}{2} NC  -\left(Ne^{-\gamma/2}\tau'\right)' \nonumber\\
	&+& \left( N^\theta p_\gamma\right)' =0\, .
\label{g12}
\ea
As needed the right sides vanish for a unique choice of lapse 
and shift and do not vanish in any other circumstance. The only solutions  
are:
\be 
N=  \frac{e^{\gamma/2}}{p} \;\;\;\;
\hbox{and} \;\;\;\; 
N^\theta=0.
\label{g14}
\ee
Hence the choice (\ref{g11})  
 selects  uniquely a Hamiltonian vector field that represents 
evolution and fixes the remainder gauge freedom. The Hamiltonian vector field 
corresponding to the lapse and shift given by (\ref{g14}) 
generates evolution along the one-parameter family of points 
labelled by $\tau(\theta)=t$. A remark is in order here. From (\ref{g12}) it is clear that the %%@
gauge choice $p_\gamma=0$ is not appropriate in this case. It would not select a hamiltonian %%@
vector field to generate evolution. Therefore from now on we restrict ourselves to the sector  in %%@
which $p\not=0$.

Finally, let us extract the true degrees of freedom of the theory.  In
order to accomplish this, we need to eliminate  
redundant variables by solving  the set of second class
constraints (\ref{g7}) and  gauge conditions (\ref{g11}). By setting $\tau
=t$ and $p_\gamma = -p$ in (\ref{g7}), we can trivially solve for
$p_\tau$ and for all but one of the infinity degrees of freedom of $\gamma$ in terms of $p$, %%@
$\psi$ and $p_\psi$ (using the
Hamiltonian and the diffeomorphism constraints respectively). The
result is:
\ba 
p_\tau & = & -\frac{t}{p}  
     \left(\frac{{p_\psi}^2} {4t^2} + \right. \left.  {\psi'}^2
     \right),
\label{g14a} \\
\gamma(\theta) & = & \frac{1}{p} \int_0^\theta d \theta_1 p_\psi \psi'+ \gamma(0)
\label{g14b} 
\ea
Let us explain the global degree of freedom left on $\gamma$. If we write %%@
$\gamma(\theta)=\frac{q}{2\pi}+\bar\gamma(\theta)$, where $q=\int_0^{2\pi}d\theta\gamma(\theta)$, %%@
to separate out the spatially variable part $\bar\gamma$ of $\gamma$, then it is clear that we %%@
can solve (\ref{g14b}) for $\bar\gamma$ and we are left with the global degree of freedom $q$ %%@
unsolved. This procedure is equivalent to Fourier analyzing $\gamma$ in space (as a function of %%@
$\theta$)  and solving for all modes but the zero mode.  
Substituting (\ref{g14b}) in (\ref{g14}), we can also express the lapse
$N$ in terms of $q$, $p$, $\psi$ and $p_\psi$. Thus, as expected, the local 
degrees of freedom reside just in the matter variables. Indeed, the
space-time metric is now completely determined by $q$, $p$, $\psi$ and $p_\psi$:
\be
g_{ab} = \frac{e^{q+\bar\gamma}}{p^2}\left(-\nabla_a t\,
\nabla_b t\, + \nabla_a \theta\nabla_b \theta \right) + t^2 \nabla_a\sigma\,
\nabla_b\sigma\, ,
\label{g14aa}
\ee
where,  from now on, $\bar\gamma$ will only serve as an abbreviation for
\be
\bar\gamma(\theta)  =  \frac{1}{p} \int_0^\theta d \theta_1 p_\psi \psi'+ \bar\gamma(0).
\label{g14aaa} 
\ee
One can show that the curvature  
invariant $R_{abcd}R^{abcd}$ for this space-time metric
blows up at $t=0$ (for $\gamma\neq const$).  
Thus, as expected, there is an initial singularity.

Recall that  the phase-space variables are periodic functions 
of $\theta$. Therefore, from (\ref{g14aaa}) we obtain a global 
constraint, that we will denote by $P_\theta$:
\be
P_\theta:= \bar\gamma(2\pi) -\bar\gamma(0)\equiv \int_0^{2\pi} 
      d \theta_1 p_\psi \psi'=0.         
\label{g14ab}
\ee
Note that this   extra global constraint arises because 
the spatial slices are compact. The presence of this extra constraint 
will make the quantum theory considerably different from the asymptotically 
flat case described in \cite{am96}.

\subsection{Reduced Phase Space}\label{rps}

The reduced phase space
$\Gamma_R$ can be coordinatized by the pairs $(q, p;\psi(\theta), p_\psi
(\theta))$. However,  the canonical variables are subject to 
the global constraint $P_\theta=0$. Therefore, the physical 
reduced phase space  is non-linear and has  the structure of a manifold 
instead of the usual vector space. Because of this 
non-linearity  it is appropriate to postpone the reduction 
by the constraint to the quantum theory. Then, it will 
be imposed as an operator condition on the quantum states.
 
 The (non-degenerate) symplectic structure on the reduced phase
space $\Gamma_R$ is the pull-back of the symplectic structure on
$\Gamma$. Thus,
\ba
\{q, p\} & = & 1
\label{g14cc}\\
\{\psi (\theta_1) , p_\psi (\theta_2) \} & = &  \delta(\theta_1 , \theta_2) 
\label{g14c}
\ea
on $\Gamma_R$. Next, let us write the reduced action by substituting
(\ref{g11}) and (\ref{g14a}) in (\ref{g5a}),
\be
S =  \int dt \left[p\dot{q} + \int_0^{2\pi} \right. 
       d\theta \left( p_\psi {\dot\psi} \right. \,  
	-\frac{t}{p}   
     \left.\left.\left(\frac{{p_\psi}^2}{4t^2} +   {\psi'}^2 \right)\right)\right],
\label{g15}
\ee
where we have been able to carry out the $\theta$ integration on the first term as a consequence %%@
of the $\theta$ independence of $p$. 
The reduced Hamiltonian, that we shall denote by $H$, is explicitly 
time dependent.  Note 
that the global constraint is conserved under time evolution because 
$\{P_\theta , H\} = 0$.
By varying
the action (\ref{g15}) with respect to $\psi$ and $p_\psi$ we obtain 
the second-order equation of motion for $\psi$:
\be
- \frac{\partial^2 \psi}{ {\partial T}^2} + 
   \frac{\partial^2 \psi}{{\partial \theta}^2} - 
   \frac{1}{T}\frac{\partial \psi}{\partial T} = 0,
\label{g17b}
\ee
where we defined a new time coordinate T on M via a constant rescaling: $T=t/p$. 
This is exactly the Klein-Gordon equation for a scalar field 
propagating on a flat background $\stackrel{\circ}{g}_{ab}$,  
given by:
\be
\stackrel{\circ}{g}_{ab} = - \nabla_a T \nabla_b T + \nabla_a \theta \nabla_b \theta  
	  + T^2 \nabla_a \sigma \nabla_b \sigma.
\label{g17c}
\ee

So we have achieved the same conceptual simplification as in the 
Einstein-Rosen model, 
i.e., the decoupling of the system.  However it is technically 
different because now  the topology of the fictitious 
background   is not ${\rm R}^3$, the flat space is not globally Minkowskian. As we will see, it 
is a wedge of Minkowski space-time identified, such that 
the boundary of the compact spatial slices is time dependent and 
starts at a singularity. 
The topology of the fictitious background and the 
geometry of the singularity on  $\stackrel{\circ}{g}_{ab}$
is more transparent if we change to a coordinate system where the 
metric is explicitly flat, i.e.,
\be
\stackrel{\circ}{g}_{ab}=
	(-\nabla_a x_0 \nabla_b x_0 +\nabla_a  \theta \nabla_b \theta)
       +  \nabla_a x_2 \nabla_b x_2.
\label{g4} 
\ee
The two sets of coordinates are related by  $x_0 = T \cosh (\sigma-\pi)$  
and $x_2=T \sinh  (\sigma-\pi)$. Let us concentrate 
on the plane defined by $x_0$ and $x_2$. Due to the angular range 
of $\sigma$, the $x_2$ coordinate will be identified, i.e., 
$x_2(0) = x_2 (2\pi)$. At $T=0$ the length of the orbit 
of $\partial/\partial\sigma$ 
goes to zero and thus the orbit reduces to a point  and the 
two-surface (torus) is a circle (on $\theta$). 
Therefore, the space-time has
 a  singularity at $x_0=T=0$. In terms of surfaces 
$T=const$, as $T$ increases the spatial slices (tori) are expanding. 

Although the fictitious metric has a singularity the initial value 
problem for the scalar field $\psi$ is well posed $\forall T>0$. The 
explicit time-dependence of the Hamiltonian reflects the 
expanding background and the existence 
of the singularity.  So, differently from the asymptotically flat case, there 
is a  (`mild') back-reaction of the gravitational field.

A remark is in order. In the  
coordinatization adopted in \cite{b74} the reduced 
system also decouples, but naively it is equivalent to 
a scalar field, denoted there by $B_- (\theta ,t)$ 
($\psi=\ln \tau -2\sqrt{3} B_-$),  
propagating on a curved background. In the light of this result, 
the fact that we got a scalar field propagating on a flat 
background might seem surprising. However,  changing the 
time coordinate $t$ via $\ln 2T=2t$ one can reinterpret the equation of 
motion of $B_- (\theta,T)$ to that of a scalar field propagating on the 
flat background $\stackrel{\circ}{g}_{ab}$.

Finally, for reasons that will be clear in the  next section, let us split the reduced phase %%@
space into its global and local degrees of freedom by defining the following direct sum: %%@
$\Gamma_R = \stackrel{\circ}{\Gamma} \oplus \bar{\Gamma}$ where $\stackrel{\circ}{\Gamma}$ is %%@
coordinatized by the canonical pair $(q, p)$and $\bar{\Gamma}$ by $(\psi (\theta), p_{\psi} %%@
(\theta)$.
Moreover let us introduce  a covariant description of the 
 phase-space $\bar\Gamma$. In this approach the  
phase-space consists of the real solutions of the 
Klein-Gordon equation  (\ref{g17b}), i.e., 

\be
 \psi(\theta,T)= \sum_{m=-\infty}^\infty
	  f_m(\theta,T) A_m +
	  f_m^*(\theta,T) A^*_m 
\label{g17d}
\ee
\ni
where  $A_m$'s are arbitrary constants and 

\ba 
f_0 (\theta,T) &=& \frac{1}{2}\left( \ln T -i\right)\nonumber \\
f_m (\theta,T) &=& \frac{1}{2}H_0^{(1)} (|m|T)e^{ im\theta} \nonumber \\ 
	       &=& \frac{1}{2}\left( J_0(|m|T) + i N_0 (|m|T)\right) e^{ im\theta}
	\quad {\hbox{for}}\quad m\neq 0
\label{g17e}
\ea
\noindent
where $H_0^{(1)}=J_0 + iN_0$ is the 0th-order Hankel function of the 1st kind 
and $J_0$ and $N_0$ are the 0th-order Bessel function of the first and 
second kind respectively. (* denotes complex conjugation.).  
 
The  symplectic structure is given by:
\be
\Omega (\psi_1, \psi_2) =  \int_{0}^{2\pi} T d\theta 
 \left( \psi_2 \partial_T \psi_1 - \psi_1 \partial_T \psi_2\right).
\label{g18}
\ee
This covariant approach is completely equivalent to the canonical 
description outlined on the first paragraph of this subsection.
We will refer to the covariant phase space as the  `space of real solutions' 
and denote by $V$.
Note that because the symplectic structure is conserved 
(${\dot\Omega}=0$), generically    the scalar field  
$\psi$ is expected to diverge at $T=0$.

\section{Quantum Theory}\label{gqt}

\subsection{Fiducial Hilbert Space}\label{gfhs}

In this subsection we will ignore the global constraint (\ref{g14ab}) 
and quantize the system.  The Hilbert space thus obtained 
will be called fiducial Hilbert space and will be denoted by ${\cal F}$.  
The physical Hilbert space is the subspace of the fiducial Hilbert 
space corresponding to the kernel of the global constraint 
and will be constructed on the next subsection. 

Recall that in addition to the local degrees of freedom $(\psi (\theta), p_{\psi}(\theta)$, which %%@
will be treated on the next paragraph as operators acting on a Hilbert space $\bar{\cal F}$, we %%@
also have the global degree of freedom ($q, p$), whose quantization is trivial. We will denote by %%@
$\stackrel{\circ}{\cal H}$ the Hilbert space where $\hat q$ and $\hat p$ are well defined %%@
operators. 
The fiducial Hilbert space, therefore, will be the tensor product of $\stackrel{\circ}{\cal H}$  %%@
and $\bar{\cal F}$, i.e., ${\cal F}= \stackrel{\circ}{\cal H}\otimes \bar{\cal F}$

We will now construct $\bar{\cal F}$. 
The overall procedure was discussed in Ref. \cite{am96}. Recall that 
in order to quantize the (unconstrained) theory  the phase space variables 
$(\psi(\theta), p_{\psi}(\theta))$ are smeared with test 
fields and represented by  operators satisfying  the commutation 
relation corresponding  to their Poisson brackets. Quantization 
is then accomplished by  constructing a Hilbert space of 
quantum states that corresponds to a *-representation of this 
observable algebra. 
The appropriate representation will be selected by imposing physical 
requirements. We will demand the time dependent  Hamiltonian  
 (\ref{g15}) and the 
global constraint  (\ref{g14ab}) to be promoted to well-defined operators.

Recall  that  the fictitious background  (M, $\stackrel{\circ}{g}_{ab}$)
is  a  suitably identified wedge of flat space-time  and that the spatial 
slices have a time dependent boundary. Moreover there is no 
global time-like Killing vector field. Therefore, 
the procedure to obtain a representation, followed in 
\cite{am96}, is not directly applicable.
We will instead adopt a   
prescription that  is outlined in \cite{am75}. We 
should add that for linear theories  both procedures 
are general and  it is just a matter of 
one being more convenient then the other  for the given problem. 

The first assumption to obtain a representation for the observable algebra 
is that, as in \cite{am96}, the Hilbert space will have  
the structure of a symmetric Fock-space $\bar{\cal F}$, based on some 
one-particle Hilbert space ${\cal H}$.  The one-particle 
Hilbert space can be constructed from the space of real solutions $V$ 
in the following way.
First one introduces on $V$ a complex structure ${\cal J}$ which is compatible 
with the symplectic  structure (\ref{g18}), i.e., 
$(\psi_1,\psi_2):=\Omega({\cal J}\psi_1, \psi_2)$ is a 
positive-definite inner product on $V$. 
Then, one can define on the complex vector space $(V, {\cal J})$ the 
inner-product:

\be
<\psi_1 | \psi_2> :=\frac{1}{2}\Omega({\cal J}\psi_1, \psi_2) +
       \frac{1}{2} i \Omega(\psi_1, \psi_2).
       \label{g18a}
\ee
\ni
Finally, the   one-particle Hilbert space   is the Cauchy completion 
of the complex inner-product space $(V, {\cal J}, <|>)$.  Furthermore, 
to complete the construction of the Fock space 
one  introduces positive and negative frequency decomposition on 
${\cal H}$ via:
\ba
 \psi^+ &:=& \frac{1}{2} ( \psi -i {\cal J} \psi)\label{g18bb}\\
 \psi^- &:=& \frac{1}{2} ( \psi +i {\cal J} \psi)\label{18b}
\ea
\ni
such  that $\psi = \psi^+ +\psi^-$. Thus, a field 
operator in the Fock space is represented by 
${\hat \psi}(\psi)= {\hat A} + {\hat C}$, 
where  ${\hat C}$ and ${\hat A}$ are creation and annihilation operators 
associated with the positive and negative frequency decomposition.

In summary,  to obtain a representation all the work can be focussed   
on finding an appropriate complex structure on $V$. 
In our model there is a natural complex structure that arises 
from the general solution (\ref{g17d}). 
Using the fact that the solution in  (\ref{g17d}) naturally occurs 
in pairs, we define a complex 
structure ${\cal J}$ for our model as:
\ba 
{\cal J} \, \alpha\ln T  := -\alpha\;\;\;\;{\hbox{and}} 
\;\;\;\; {\cal J} \,\alpha := \alpha\ln T
\;\;\;\;&{\hbox{for}}& m = 0 \label{g20a} \\
{\cal J} \, J_0(|m| T) := N_0 (|m| T)\;\;\;\;{\hbox{and}} 
\;\;\; {\cal J} \, N_0(|m| T) := - J_0(|m| T) 
\;\;\;&{\hbox{for}}& m\neq 0 
\label{g20}
\ea
($\alpha$ is a constant.) 
Note that $J^2=-1$ as required. 
Thus, the ``positive frequency'' part  
can be obtained by using (\ref{g17d}), (\ref{g18bb}), 
(\ref{g20a}) and (\ref{g20}): 
\be
\psi^+=\sum_{m=-\infty}^\infty f_m^*(\theta,T)A_m^*,
\label{20aa}
\ee
\noindent
and, from (\ref{18b}), the negative frequency part is just the complex 
conjugate of (\ref{20aa}).
Moreover one can show that the 
complex-structure, given by (\ref{g20a}, \ref{g20}), is in fact 
compatible with the symplectic structure  
and therefore completely determines  
the inner-product on the one particle Hilbert space ${\cal H}$.  
Following the prescription we can write the field operator 
${\hat \psi}$ in terms of  creation and annihilation 
operators corresponding   to  the positive and negative   frequency
decomposition defined on ${\cal H}$ by the complex structure 
(\ref{g20a}, \ref{g20}): 
\be
{\hat \psi}(\theta,T)= \sum_{m=-\infty}^\infty 
	  \left( f_m(\theta,T){\hat A}_m +  \right.
	  \left.f_m^*(\theta,T){\hat A}^{\dagger}_m  \right)
\label{g19}
\ee
A comparison with the classical solution  
(\ref{g17d}) shows that one could obtain this field operator naively 
by promoting the constants of motion from the explicit  solution  
to  operators. In fact this is also a  motivation  for 
choosing (\ref{g20a}, \ref{g20}) as   the complex structure.  
Furthermore, as needed, one can show that the normal ordered 
Hamiltonian and global constraint are well-defined operators.

There is an important result from the above construction.
Note that the complex-structure is time {\it independent},  therefore, 
there is no mixing between positive 
and negative parts or analogously, creation and 
annihilation operators as time passes.  
Thus, in contrast to ref.\cite{b74}, we conclude that there is no-creation 
of `particles'.

\subsection{Physical Hilbert Space}

The space ${\cal F}_{\sl phys}$ of physical states   is the subspace 
of the fiducial Hilbert space $\cal F$ defined by:

\be
:{\hat P}_\theta:|\Psi>_{\sl phys}=0. 
\label{g20b}
\ee
In order  
to have a better understanding of this condition let us 
express the global constraint in terms of creation and annihilation operators: 

\be
:{\hat P}_\theta:|\Psi>_{\sl phys}=2{\sum_{m=-\infty}^\infty}
   m {\hat A}^{\dagger}_m {\hat A}_m\, |\Psi>_{\sl phys}\,=\,  0.
\label{g21}
\ee
\ni
Thus, the physical states are  states such that   the
total angular momentum in $\theta$-direction vanishes.  
Therefore, obviously 
the (usual) vacuum state  defined by ${\hat A}_n\otimes {\hat q}|0>=0$, 
$\forall\; k$ and states with particles in the zero mode (i.e. $m=0$) 
belong to the physical Hilbert space. 
Explicitly, a generic physical 
state with $N$ particles is given by:

\be
|{}^N\Psi>_{\sl phys} = |\phi> \otimes\left[\prod_{i=1}^N |m_i>\right] \quad{\hbox{such %%@
that}}\quad
      \sum_{i=1}^{N} m_i =0.
\label{g22}
\ee
\ni
where  $|\phi>$ is a state that belongs to $\stackrel{\circ}{\cal H}$ and $|m_i>$ represents a %%@
one-particle state with angular momentum $m_i$.
Note that, the space of physical states does not inherit the  Fock space 
structure of ${\bar{\cal F}}$.  
For instance, except for the zero mode,  none of the one-particle 
states of $\bar{\cal F}$  
belongs to  ${\cal F}_{\sl phys}$.  
Nonetheless  because the orbit of $P_\theta$ is compact (or 
$:{\hat P}_\theta:$ has discrete spectrum)
${\cal F}_{\sl phys}$ is a closed subspace of ${\cal F}$ and 
hence has a natural Hilbert space structure.  Moreover, 
 the operators from 
${\cal F}$ can be projected to ${\cal F}_{\sl phys}$.  We will 
denote the projection operator  by ${\cal P}$.

Let us now investigate the dynamics of the system.  The 
Hamiltonian, from expression (\ref{g15}), can be  promoted to the 
operator:
\be
{\hat H}(T) = \int_0^{2\pi}d\theta 
      :\hat{p}T\left(\frac{{\hat p}^2_\psi}{4\hat{p}^2T^2} +   ({\hat \psi}')^2 \right):,
\label{g23}
\ee
\ni
where we have rescaled the time as before. Expressing in terms of creation and annihilation %%@
operators, we obtain:
\ba 
{\hat H} &=& \frac{\pi}{2T}:\hat{p}({\hat A}_0 + {\hat A}^\dagger_0)^2: \nonumber\\
  &+& \frac{T}{4} \sum_{m=-\infty}^{\infty} m^2  \hat{p}          
    \left[ 2 {\hat A}^{\dagger}_m {\hat A}_m (H_1^{(1)}H_1^{(2)}+ %%@
H_0^{(1)}H_0^{(2)})\right.\nonumber\\
  &+& {\hat A}_m{\hat A}_{-m} \left( (H_0^{(1)})^2+  (H_1^{(1)})^2\right)
    + \left.{\hat A}^{\dagger}_m {\hat A}^{\dagger}_{-m} \left( (H_0^{(2)})^2+ %%@
(H_1^{(2)})^2\right)\right],
\label{g24}
\ea
where $H_0^{(2)}(|m|T)=[H_0^{(1)}(|m|T)]^*$ and 
$H_1^{(1,2)}(|m|T)=-\frac{1}{|m|}{\dot H}_0^{(1,2)}(|m|T)$.   
By inspection  one can easily conclude that the action of 
the Hamiltonian operator on a physical state gives back a 
physical state, thus it leaves ${\cal F}_{\sl phys}$ invariant.   
Note that the vacuum state is {\it not} an eigenvector of the Hamiltonian 
with zero eingenvalue. 
However, this Hamiltonian arises from 
our choice of deparametrization, therefore apriori there is no 
direct physical significance to this result. In contrast to the Einstein-Rosen model, 
where the vacuum was physically defined, here it is not clear how to 
define the `real', physical  vacuum.  This is a consequence of the 
arbitrariness of the deparametrization procedure.

In order to investigate the time parameter that arises from the 
quantum theory let us write the Schr{\"{o}}dinger equation 
associated with  (\ref{g23}):

\be
i\hbar \frac{\partial}{\partial T}|\Psi>_{\sl phys}(T) = {\hat H}(T) 
	     |\Psi>_{\sl phys}(T).
\label{g24a}
\ee
\noindent
One can interpret this equation by saying that this 
evolution takes place on the fictitious background  
$\stackrel{\circ}{g}_{ab}$ where 
$\frac{\partial}{\partial T}$ has a space-time interpretation.   
The physical space-time is a derived quantity, i.e., 
there is no physical metric to start with, thus apriori 
$\frac{\partial}{\partial T}$ has no physical 
interpretation. Nonetheless, 
because of the classical decoupling we can still write 
the metric operator by using the chart $(T,\theta,\sigma)$. 
This is remarkable because  we can conclude that 
in fact, although there is no background physical metric, 
the simplification 
of this system is  such that provides a time parameter 
for the quantum theory.

\subsection{Quantum Geometry}\label{gqg}

Let us start investigating semi-classical geometries. 
In order to do this,   
we shall return to the fiducial  Hilbert space ${\cal F}$. 
First we promote the classical 
expression of the space-time 
metric (\ref{g14aa})  to an operator on ${\cal F}$. Formally,
\be
{\hbox{``}}\; {\hat g}_{ab} =  :e^{\hat{q}+{\hat {\bar\gamma}}(\theta,T) }:\left(-\nabla_a T\,
\nabla_b T\, + \nabla_a \theta\nabla_b \theta \right) + \hat{p}^2 T^2 \nabla_a\sigma\,
\nabla_b\sigma\;{\hbox{''}}\, .
\label{g29}
\ee
\noindent
As in the   Einstein-Rosen case, the states in $\bar{\cal F}$ yielding semi-classical 
geometry will be the usual coherent states for the field 
operators. Similarly,  the matrix elements of 
this operator on coherent states on $\bar{\cal F}$ are  
well-defined.  In particular, the expectation value on a coherent state 
$|\Psi_c>$ in ${\cal F}$, gives the classical expression (\ref{g14aa})(assuming the time %%@
rescaling)  evaluated on 
the  classical field configuration $\psi_c$ and the mean values $<{\hat q}>$ and $<{\hat p}^2>$ %%@
peaked on  a classical solution $q$ and $p^2$, explicitly

\be
<\Psi_c|{\hat g}_{ab}|\Psi_c> =  
   e^{q+ 2\int_0^{\theta} 
 T {\dot  \psi_c} \psi_c' d\theta_1+ {\bar\gamma}(0)}
 \left(-\nabla_a T\,\nabla_b T\, + \nabla_a \theta\nabla_b \theta \right) 
 + p^2 T^2 \nabla_a\sigma\, \nabla_b\sigma\, .
\label{g29a}
\ee
\noindent
An immediate consequence of this result is that the coherent state 
provides us an example to show that the singularity persists 
in the quantum theory.  Specifically, we can compute a  
scalar formed out  from the Riemann tensor, promote to a (normal ordered) 
quantum operator and calculate its expectation value on a coherent 
state. We will, then,  obtain the corresponding divergent classical value.

The only non-trivial metric operator component  is  
 $e^{\hat{q}}  \exp[(\int_0^\theta 2T {\dot{\hat \psi}}{\hat 
\psi'}d\theta_1)+{\hat {\bar\gamma}}(0)]$. Note that  the exponent of the second factor  has the same functional form as  
the angular momentum for a scalar 
field in a box (if one recovers the integral on $\sigma$ that is 
omitted due to symmetry). Similarly to the energy 
in a box that was  extensively discussed in Ref. \cite{am96} and \cite{m00}, 
one can show that it 
is not a well-defined operator on $\bar{\cal F}$. Thus, after regularizing  we  
obtain the regulated metric operator on $\bar{\cal F}$:
\be
 {\hat g}_{ab}(f_\theta) =  e^{\hat{q}}e^{ \int_0^{2\pi} 
 f_\theta(\theta_1)2T :{\dot{\hat \psi}}{\hat \psi'}:d\theta_1+{\hat {\bar\gamma}}(0)}
 \left(-\nabla_a T\,\nabla_b T\, + \nabla_a \theta\nabla_b \theta \right) 
 + {\hat{p}^2}T^2 \nabla_a\sigma\, \nabla_b\sigma\, ,
\label{g30}
\ee
where the regulator $f_\theta(\theta_1)$  equals $1$ for 
$\theta_1\leq \theta- \epsilon$, then it smoothly decreases to zero 
and equals zero for $\theta_1 \geq \theta-\epsilon$. Also,  
$f_{2\pi}(\theta)=f_0(\theta)$. 
As we see, again 
the Planck length comes into play naturally in the quantum theory

Although we have obtained a well-defined operator on $\bar{\cal F}$, 
it is not an operator on the physical space because it 
does not commute with the global constraint $:{\hat P}_\theta:$.  
However, as we pointed out before, it can be projected to the 
physical space.  Therefore the physical regulated metric operator 
in ${\cal F}_{\sl phys}$ is given by: 
\be
[{\hat g}_{ab}(f_\theta)]_{\sl phys}= {\cal P} \,{\hat g}_{ab}(f_\theta)\,{\cal P}.
\label{g31}
\ee
\noindent

As in the Einstein-Rosen waves, there are interesting 
 features associated to the metric operator. Because 
the calculations are similar, we will avoid the 
repetition by pointing out only the final results. First there  
will be quantum fluctuations of the light-cone. Specifically, 
for a given quantum state of the system (gravity coupled to scalar field), 
the norm of a null vector will fluctuate between positive, 
null and negative values.   Second the commutator between two 
non-trivial metric operators   is non-vanishing, this is 
a consequence,  as before, of the non locality of the metric 
components with respect to the (basic) scalar field. Finally, 
the holonomy operator is well-defined in this model as well. Regarding 
this operator there are some differences that are worth 
 pointing out in more detail.  Recall that in the Einstein-Rosen waves 
the first internal gauge choice that we made led to  a badly-behaved 
connection at the origin (Eq. (24) in Ref. \cite{am96}). Therefore, a gauge 
transformation   
was necessary. After that, the component of the connection along 
$\nabla_a R$ was not Abelian anymore, therefore the holonomy  
along a loop $\sigma=const$ was not trivial to compute. It was 
given by the path-ordered exponential. Thus, we explicitly 
computed only the holonomy along a loop defined by $R=const$ 
 because in this case the   component of 
the connection that was contributing was Abelian.  However, 
now using the same (first) internal gauge choice,  
we obtain a well-defined connection 
because of the topology of the spatial slices.  Moreover, the holonomy 
along both generators of the torus can be easily computed, 
because the respective contributing components of the 
connection are Abelian.  The  corresponding operators
are given by:
\be
{\hat T}^0_\eta = 2\cosh\left[\pi e^{-:{\hat \gamma}(f_\theta):/2}\right]:,
\label{ghol1}
\ee
\noindent
for a loop $\eta$ with tangent vector given by ${\dot \eta}^a=\sigma^a$ 
(along the integral curve of the Killing field),   
and 
\be
{\hat T}^0_\eta = 2\cosh\left[\frac{{\hat H}}{\hat{p}}\right],
\label{ghol2}
\ee
where now the loop  $\eta$ has tangent vector given by 
${\dot \eta}^a=\theta^a$ and ${\hat H}$ is the Hamiltonian operator   
(\ref{g23}).   Note that in the Einstein-Rosen case 
we calculated the holonomy for a loop along the integral curve 
of the Killing field  and it needed to be regulated. Here,  the  
operator (\ref{ghol1}) has to be regulated as well to yield
a well-defined operator on $\bar{\cal F}$. Moreover it does not  
commute with the global constraint, therefore it has to be 
projected to ${\cal F}_{\sl phys}$. 
On the other hand, the holonomy operator (\ref{ghol2}) does 
not require any regularization procedure (other than normal ordering).  
Moreover, it commutes with the 
global constraint. Therefore it is 
automatically a well-defined operator on ${\cal F}_{\sl phys}$,  
as expected (see discussion in Ref. \cite{am96}).

\subsection{Physical Coherent States}\label{pch}

In this section we will obtain the physical coherent states. 
In the fiducial Hilbert space, the  coherent 
states  were exactly the 
usual coherent states for the field operator in $\bar{\cal F}$ tensor product with a quasi-%%@
classical state in $\stackrel{\circ}{\cal H}$. 
However, in general, the field operator 
${\hat \psi}(\theta,T)$,  
or equivalently the  annihilation and creation operators, 
${\hat A}_m$ and ${\hat A}^{\dagger}_m$ (for $m\neq 0$), 
do not  leave   the physical Hilbert space invariant. 
Therefore, to obtain a coherent state   
on  ${\cal F}_{\sl phys}$ we will adopt the following strategy:
First we will obtain a set of `basic' operators such that any  
operator on  ${\bar{\cal F}}_{\sl phys}$  can be expressed as 
a combination of them. They play a role on  ${\bar{\cal F}}_{\sl phys}$ 
analogous  to ${\hat A}_m$ on $\bar{\cal F}$. Then, 
having obtained these operators, we will seek  states where 
they will be  peaked on their classical value. These 
will be the coherent state on   ${\bar{\cal F}}_{\sl phys}$
because by construction all other normal ordered operators will be 
peaked on their classical value as well. Finally we will obtain the coherent state on 
${\cal F}_{\sl phys}={\stackrel{\circ}{\cal H}} \otimes {\bar{\cal F}}_{\sl phys}$ in the obvious %%@
way. 

Let us obtain the set of `basic' operators on   ${\bar{\cal F}}_{\sl phys}$.  
Note, first,  that classically the phase space ${\bar\Gamma}$
modulo the  global constraint $P_\theta$ can be coordinatized by 
the infinite set:
\be
 \beta_{ \{ m_i\} }^{[N]} = \prod_{i=1}^N  A_{m_i}\quad{\hbox{and}}\quad
      \left[  \beta_{ \{ m_i \} }^{[N]} \right]^* = \prod_{i=1}^N  A_{m_i}^*
      \quad{\hbox{with}} 
      \quad \sum_{i=1}^N m_i =0 
\label{g24b}
\ee
\ni
(The infinite set $A_m$, $A_m^*$ is subject
to {\it one} global constraint, thus we are still left with an infinite set.)
Now, the `basic' set of operators can be obtained  
by promoting  (\ref{g24b})  to operators on  ${\bar{\cal F}}_{\sl phys}$:

\ba
{\hat \beta}_{ \{ m_i\} }^{[N]} &=& \prod_{i=1}^N {\hat A}_{m_i}\quad{\hbox{with}}
     \quad \sum_{i=1}^N m_i =0 \nonumber\\
\left[ {\hat \beta}_{ \{ m_i \} }^{[N]} \right]^\dagger &=& \prod_{i=1}^N {\hat %%@
A}_{m_i}^\dagger\quad{\hbox{with}} 
      \quad \sum_{i=1}^N m_i =0 
\label{g25}
\ea

We can now ask if there exists a  state on 
${\bar{\cal F}}_{\sl phys}$ with the following property,

\be
{\hat \beta}_{\{m_i\}}^{[N]}|{\bf \beta}_c> = \beta_{\{m_i\}}^{[N]} |{\bf \beta}_c>.
\label{g26}
\ee
\ni
This will be the coherent state that we are looking for. 
The answer is that there exists such state and it is given 
explicitly  by:

\ba
|{\bf \beta}_c> &=& |0> + \beta_0^{[1]}|1(0)> + 
 \sum_{m=-\infty}^\infty  
\beta_{\{m\}}^{[2]}|1(m),1(-m)> +\cdots \nonumber\\
&+& \cdots +\sum_{\{J\}}
\frac{\beta_{\{J\}}^{[N]}}{\sqrt{N_1!} \cdots\sqrt{N_I!}}
|N_1(1),\cdots,N_I(I),\cdots >+  \cdots
\label{g27}
\ea
\ni
where the sum stands for all possible sets $\{J\}$ such that  
\be
\sum_{J=-\infty}^\infty J N_J=0\quad{\hbox{and}}\quad 
\sum_{J=-\infty}^\infty N_J=N\,.
\ee
\noindent
We have changed the notation slightly in order to adopt 
the basis of the number operator.  To make the notation clear, 
let us give an example of two possible sets of this sum with three 
particles, i.e., $N=3$: 
\be
\frac{1}{\sqrt{2}}\beta_{(1,1,-2)}^{[3]}|2(1),1(-2)> 
\quad {\hbox{and}}\quad \beta_{(12,-4,-8)}^{[3]}|1(12),1(-4),1(-8)>.
\ee

Finally, a coherent state $|\Psi_c>_{\sl phys}$ on ${\cal F}_{\sl phys}$ will be given by %%@
$|\Psi_c>_{\sl phys}=|{\bf \beta}_c>\otimes |\phi_c>$, where $|\phi_c>$ belongs to %%@
${\stackrel{\circ}{\cal H}}$ and is such that as before ${\hat q}$ and ${\hat p}$ are peaked on a %%@
classical solution given by $q$ and $p$. 

The expectation value of the physical metric operator given by (\ref{g30})  
on $|{\bf \beta}_c>$ yields by construction:
\be
{}_{\sl phys}<\Psi_c|[{\hat g}_{ab}]_{\sl phys}|\Psi_c>_{\sl phys} = [g_{ab}(q, p, %%@
\beta_c,\beta_c^*)]_{\sl phys},
\label{g27a}
\ee
\noindent
i.e., classical geometries on the phase space with the global constraint 
implemented.  
Note that, as before, there is no need to have the regulator in this case.

Another interesting result is that 
one can show  that the physical coherent state  is the projection 
to ${\cal F}_{\sl phys}$ of the 
 coherent state ($<\Psi_c|{\hat \psi}|\Psi_c>=\psi_c$)  
on $\bar{\cal F}$, i.e.,

\be
|\Psi_c>_{\sl phys} = {\cal P}|\Psi_c>.
\label{g28}
\ee
\noindent

In particular, note that if $\psi_c$ is a classical solution 
corresponding to the zero mode, that we will denote by ${}^0\psi_c$, 
then $|{}^0\Psi_c>_{\sl phys}=|{}^0\Psi_c>$, ie., the physical coherent 
state for this classical solution is equivalent to the (usual) 
coherent state. 

As a consequence of (\ref{g28}) the expectation value of any 
operator projected 
to ${\cal F}_{\sl phys}$ on a physical coherent state  has the 
following correspondence with respect to the operator  and  
 coherent states on    ${\cal F}$:

\be
{}_{\sl phys}<\Psi_c |{\hat O}_{\sl phys}| \Psi_c >_{\sl phys} = 
       <\Psi_c|{\hat O}_{\sl phys}|\Psi_c>=
       {}_{\sl phys}<\Psi_c|{\hat O}| \Psi_c>_{\sl phys}.
\label{g28a}
\ee

\section{Discussion}
\label{gdiscussion}

Our choice of midi-superspace variables was motivated by our previous
work on Einstein-Rosen waves \cite{am96} and differs from the
literature on cosmological models. Our intention was to obtain the
same type of simplifications as in the Eisntein-Rosen case and this
was indeed achieved.  We saw that although this model had formally the
same bulk Lagrangian density as the Einstein-Rosen waves (with $\tau$
replaced by $R$), because the spatial slices are compact the action is
different; it does not have a boundary term.  In the absence of a
non-vanishing Hamiltonian we need to deparametrize the theory. We
chose $\tau(\theta)$ as the time coordinate.  In the Einstein-Rosen
case $R(r)$ was chosen as the radial coordinate. This difference has a
counterpart in space-time language: In the Gowdy models, due to
spatial compactness, the gradient of $\tau(\theta)$ is time-like
everywhere on $M$ whereas in the Einstein-Rosen waves $\nabla_a R (r)$
is space-like.  As is usual in the deparametrization approach, the
momentum canonically conjugate to $\tau(\theta)$ turned out to be the
Hamiltonian of the system.  In the Einstein-Rosen waves, by contrast,
$p_R(r)$ does not play a special role.

The reduced system was again remarkably simple. However, there are
some fundamental differences from the non-compact case. The infinite
number of true degrees of freedom are represented again by the scalar
field but now there is in addition one global degree of freedom. The
reduced Hamiltonian is explicitly time-dependent.  There exists a new
global constraint that requires the total angular momentum in
$\theta$-direction to vanish.  We decided to carry it over as an
operator condition on the quantum theory. Next, the space-time now has
an initial singularity.  Finally, although the decoupling occurs as in
the Einstein-Rosen case, now the scalar field propagates on a wedge of
flat space-time with certain identifications, rather than on a full
Minkowski space-time. An analysis of this identified background
space-time showed that the spatial slices are tori with time-expanding
boundary that start at a singular circle.

Because of the topological differences of the fictitious flat
background and the absence of a time-like Killing vector field, the
procedure to obtain the quantum representation is different. Now, the
representation is directly related to the choice of a complex
structure.  This approach is more suitable for linear theories and
does not rely on the staticity of the background whereas the approach
used in \cite{am96} (to find a suitable measure on the quantum
configuration space) is more convenient if there is a static Killing
vector field (but can be extended also to non-linear theories).  The
complex structure for our model was chosen accordingly to physical
requirements. This led us to a fiducial Hilbert space.  An interesting
problem is to work out the corresponding measure in the space of
quantum states $L^2(S',d\mu)$.

In contrast to ref. \cite{b74} , we obtained one fixed Hilbert space,
rather than a one-parameter family of them. This is a consequence of
the fact that our complex structure is time
\textit{independent}. Another implication is that there is no creation
of `particles'.  Indeed, one could obtain particle creation even in
Minkowski space-time just by a choice of time dependent complex
structure.  Here we do not have external fields, i.e., we are not
quantizing fields in a curved space-time.  Therefore, there is
no reason to expect creation of particles or even to interpret the
quanta of the scalar field as physical particles.

The physical Hilbert space is the subspace of the fiducial state space
defined by the kernel of the global constraint. There are no
one-particle physical states, except for the zero mode.  We saw that
because the orbits of the classical constraint vector field are
closed, the quantum physical operators can be obtained by simple
projection into the physical subspace.  We also obtained
semi-classical states.  The expectation value of the metric operator
in such a state corresponds to classical geometry.

The issue of time in this model is technically simpler but
conceptually more subtle. As in the Einstein-Rosen case the gauge
choice is such that the problem decouples and a time parameter arises
naturally in the quantum theory. It is the time parameter of the
fictitious background.  The conceptual problem is that while we had a
distinction between gauge and dynamics and a notion of time at spatial
infinity before, now the time parameter and the Hamiltonian are
artifacts of our deparametrization, i.e., they are not singled out by
any physical reason. But, as is well known, this problem is intrinsic
to the theory itself and is also present in the classical theory.  The
key point is that we were able to find a {\it gauge} which selects a
fictitious background and provides a global time parameter for the
quantum theory as well.

In this model, as we pointed out, there is an initial singularity in
all classical solutions, including the flat fictitious background.
Nonetheless we were able to quantize the system consistently. Note,
however, that the quantization did not cure this break down of the
classical theory. However, since the symmetry reductions freeze out
degrees of freedom, and hence it is not clear that the problem will
persist in the full quantum theory.

A supportive result for the non-perturbative program based on
Ashtekar's self-dual connections is that the operator corresponding to
the trace of the holonomy of a loop around the axis of symmetry is
well-defined on the Hilbert space of both models.  The holonomy along
a loop in the perpendicular direction, due to axi-symmetry,
corresponds to a 2-dimensional smearing.  Therefore, it is expected to
be well-defined. We were able to compute this holonomy as a simple
exponential, and we verified that, in fact, the operator corresponding
to its trace is well-defined without need of regularization (other
than the usual normal ordering).  It is not clear, however, how the
first result will change if we adopt a different gauge choice and it
may well be that a smearing of the loop would be necessary then. Here
we saw that there is no need of such smearing because the metric is a
surface integral of the basic fields.

To summarize, this model can be successfully used to probe a number of
features of the quantum theory.  The key reason is that it is a
simple, exactly soluble model.  It makes potential problems
transparent and suggest methods to deal with them.

\vskip.5cm
\ni
{\Large\bf  Acknowledgments}

I would like to thank A.Ashtekar for suggesting this problem and advising through out its completion. I would also like to thank J.Pullin and C. Cutler for discussions and G. Mena Marug{\'a}n for comments  and suggestions on the first draft of this paper.

\end{document}